# Origin of high-frequency radiation during laboratory earthquakes.

**S. Marty[1*], F. X. Passelègue[2], J. Aubry[1], H. S. Bhat[1], A. Schubnel[1] and R. Madariaga[1]**

[1]Laboratoire de Géologie, École Normale Supérieure/CNRS UMR 8538, PSL Research

University, 24 rue Lhomond, F-75005 Paris, France.

[2]École Polytechnique Fédérale de Lausanne, CH-1015 Lausanne, Switzerland.

*Corresponding author: Samson Marty (marty@geologie.ens.fr)

**Key points**

- High-frequency radiation is enhanced with both confining pressure and rupture velocity
- Acoustic sensors can be used as an array to track high-frequency sources during rupture propagation
- High-frequency radiation sources propagate consistently with the rupture front and are located behind it

**Abstract**

We monitor dynamic rupture propagation during laboratory stick-slip experiments performed on saw-cut Westerly granite under upper crustal conditions (10-90 MPa). Spectral analysis of high-frequency acoustic waveforms provided evidence that energy radiation is enhanced with stress conditions and rupture velocity. Using acoustic recordings bandpass filtered to 400-800 kHz (7-14 mm wavelength) and highpass filtered above 800 kHz, we back projected high-frequency energy generated during rupture propagation. Our results show that the high-frequency radiation originates behind the rupture front during propagation and propagates at a speed close to that obtained by our rupture velocity inversion. From scaling arguments, we suggest that the origin of high-frequency radiation lies in the fast dynamic stress-drop in the breakdown zone together with off-fault co-seismic damage propagating behind the rupture tip. The application of the back-projection method at the laboratory scale provides new ways to locally investigate physical mechanisms that control high-frequency radiation.

**Plain Language Summary**

Over geological timescales, partially or fully locked tectonic plates accumulate stress and strain. The stress and the strain build-up on discontinuities that we call "faults". Natural faults exist either inside a tectonic plate or at the boundary between two distinct tectonic plates. When the stress accumulated on a fault exceeds the strength of the fault, the accumulated stress and strain, which can be interpreted in term of accumulated energy, are suddenly released. This natural phenomenon is called an "earthquake". During an earthquake, part of the energy is released in the form of seismic waves. Those seismic waves are responsible for the ground shaking. High-frequency waves usually cause most of the damage. To better understand the physical parameters that influence the generation of high-frequency waves, we experimentally reproduced micro-earthquakes and used them as a proxy to study real earthquakes. Our results showed that the higher the pressure acting on the fault when an earthquake is generated, the higher the amount of high-frequency radiations. Moreover, our observations underlined that, during an earthquake, high-frequency waves are released in specific areas on the fault. Thus, these results might be of relevance to improve seismic hazard assessment.

**1 Introduction**

Even though high-frequency waves (> 1 Hz) are likely to be the most damaging during earthquakes propagation, physical processes at the origin of high-frequency radiation are still under debate and relatively less well understood (Das, 2007). First kinematic models used to invert seismic slip distribution (Haskell, 1964, Savage, 1966) were unable to describe high-frequency radiation because they assumed flat source models with constant slip and stress drop on the fault.

Fracture models which introduced variable slip function and rupture velocity showed that changes in rise time and rupture velocity lead to high-frequency radiation (Madariaga, 1977, Madariaga, 1983). Later, seismologists used ray-theory to calculate high-frequency radiation from earthquakes having spatial variations of rupture velocity, slip velocity and stress drop (Bernard and Madariaga, 1984, Spudich and Frazer, 1984) and predicted thatthe starting and stopping phases of earthquakes to be responsible of high-frequency radiation. A good illustration of this phenomena is the January 17th 1984 Northridge earthquake (Mw 6.7) for which Hartzell at al [1996] identified the initiation of the rupture and its stopping to be concurrent with high-frequency radiation.

An interesting case of the rupture velocity effects on high-frequency radiation is that of earthquakes propagating at supershear velocities (i.e. velocities higher than the shear wave speed). Supershear earthquakes are suspected to be more devastating than sub-Rayleigh earthquakes (with rupture velocities slower than the S-wave velocity) due to the formation of Mach-wave fronts (Dunham, and Archuleta, 2004, Bhat et al., 2007, Bruhat et al., 2016). Theoretical studies of supershear rupture (Hamano, 1974, Andrews, 1976, Das and Aki, 1977) followed by experimental works on plastic polymer (Wu et al, 1972, Rosakis et al., 1999) demonstrated the existence of possible supershear scenarios. Following the Mw 7.6 devastating Izmit earthquake in Turkey, Bouchon et al. [2001] successfully made the observation that certain parts of the fault ruptured at supershear speeds. Passelègue et al. [2013] were the first to experimentally illustrate the rupture transition from sub-Rayleigh regime to supershear regime on centimetric rock samples at upper crustal stress conditions. In these experiments, Passelègue et al. [2016] observed particularly energetic high-frequency radiation during stick-slip rupture propagation, the origin of which remained obscure.

Quite recently, the emergence of dense and large aperture seismic arrays has provided a new method to investigate the spatial and temporal behavior of seismic energy release during large earthquakes. This method, called back-projection, utilizes the time-reversal property of seismic waves to retrieve their sources and was introduced by Spudich and Frazer [1984]. Following the successful application of the back-projection method to the 2004 Sumatra-Andaman earthquake by Ishii et al. [2005], the back-projection method has been applied to numerous earthquakes (Kiser et al., 2011, Okuwaki et al., 2014, Zhang and Ge, 2010, Ishii, 2011, Wang and Mori, 2011). To the best of our knowledge, the technique has never been applied in the laboratory yet, where it might shed light on the origin of high-frequency radiation.

This study presents results from stick-slip experiments conducted on saw cut Westerly granite under tri-axial conditions and is devoted to investigate the dynamics of high-frequency radiation during rupture propagation. First, the rupture velocity of dynamic stick-slip instabilities was measured using piezoelectric acoustic sensors by tracking the propagation of the rupture front. We then investigate the influence of stress conditions and rupture velocity on high-frequency radiation. Second, we apply the back-projection method to image high-frequency sources during rupture history and discuss their link to rupture front propagation.

## 2 Experimental set-up

Stick-slip experiments were performed on Westerly granite using a tri-axial oil-medium loading cell ($\sigma_1 > \sigma_2 = \sigma_3$). The confining pressure and the differential stress (i.e the axial stress) can go up to 100 MPa (about 3km depth) and 700 MPa respectively. Experiments were conducted on Westerly granite which is a rock-mechanics standard with millimetric grain sizes and P, S and Rayleigh wave velocities that are respectively 5700, 3500 and 3200 m/s (Scholz, 1986). Cylindrical samples were 40 mm wide and 88 mm long and were cut at an angle of 60 degrees from the horizontal plane in order to create a weak fault interface. Fault interface was roughened with a #160 grit paper to create homogeneous roughness and to minimize cohesion. The axial displacement of the piston, the confining pressure and the axial stress were measured by external sensors.

Acoustic emissions were recorded during the experiments using a high-frequency acoustic monitoring system at a sampling rate of 10 MHz. There were 16 piezo-ceramics acoustic sensors that were used in this study. All the acoustic sensors were polarized in the same way and were mostly sensitive to P-waves (i.e. motion perpendicular to the sample surface). A complete description of the tri-axial apparatus and of the high-frequency acoustic monitoring system is given in the supplementary material and in Passelègue et al. (2016).

## 3 Methodology

In our study we subdivide the 16 acoustic sensors into two arrays. The first array consists of 7 acoustic sensors evenly distributed along the fault plane which were used to monitor the rupture front propagation. The 9 remaining acoustic sensors form the second array, which is used to both locate the nucleation zone of the stick-slip instability and for the back-projection analysis. The 9 sensors were arranged as close as possible to each other and face the fault. Hereafter, we refer to the first array as AFAS (along fault acoustic sensors) and to the second array as OFAS (off-fault acoustic sensors). The geometry of both arrays is shown in the supplementary material (Figures S1b and S1c).

### 3.1 Rupture velocity inversion

Previous studies have already used acoustic sensors to monitor rupture front propagation during stick-slip instability either on plastic polymers (Schubnel et al., 2011) or crustal rocks (Passelègue et al., 2013). Linear elasticity predicts the existence of an elastic strain singularity at the head of the rupture tip which is proportional to $r^{-n}$ where $r$ is the distance to the rupture tip and $n$ an exponent which depends on the rupture velocity ($0 <= n <= 0.5$). Acoustic sensors located along the fault will record the passage of the rupture front and can be used to estimate the rupture velocity.

In our rupture velocity inversion we apply the following methodology: (i) P-wave arrival times are manually picked on OFAS recordings and are used to determine the initiation time as well as the location of the nucleation zone on the fault (ii) using the least square method, we search for the average rupture velocity that best matches the observed rupture front arrival times on the AFAS recordings. The method is exhaustively described in the supplementary material and in Passelègue et al. (2013, 2016).

### 3.2 The back-projection method

The back-projection technique propagates seismogram waveforms backward in time to a grid of potential sources, in order to determine the spatial and temporal evolution of seismic sources during an earthquake. The strength of the technique lies in its simplicity since it only requires a velocity structure model and a grid of potential sources.

In the present study, we use the coherency function $x(t)$ first introduced by Ishii et al. [2011] to track high-frequency sources during rupture propagation. The coherency function quantifies the average cross-correlation over a time window $T$ of the stacked waveform and each individual acoustic waveform. For a set of $k$ acoustic sensors, at a time $t$ and from a source $i$, the coherency function $x_i(t)$ takes the form:

$$x_i(t) = \frac{1}{k}\sum_{n=1}^{k}\frac{p_n\sum_{\tau=t}^{t+T}u_n(\tau+t_{i,n}+\Delta t_n)*s_i(\tau)}{\sqrt{\sum_{\tau=t}^{t+T}u_n^2(\tau+t_{i,n}+\Delta t_n)}\sqrt{\sum_{\tau=t}^{t+T}s_i^2(\tau)}}$$

where $s$ is the stacked waveform which for a source $i$ and at time $t$ takes the form :

$$s_i(t) = \frac{1}{k}\sum_{n=1}^{k}w_n u_n(t+t_{i,n}+\Delta t_n)$$

with $k$ the total number of acoustic sensors, $u_n(t)$ the recorded acoustic waveform of the $n$'th acoustic sensor, $t_{i,n}$ the predicted P-wave travel time between $i$'th grid location and the acoustic sensor $k$, $\Delta t_n$ the time correction of the $n$'th acoustic sensor that we obtain by cross-correlating the initial few micro-seconds of each acoustic waveform with a reference waveform. $\Delta t_n$ ensures that all waveforms align well at the nucleation location. The cross-correlation also yields the weighting factor $w_n = p_n/A_n$ with $p_n$ that corrects for first P-wave polarity (either equals to -1 or 1) and $A_n$ a normalization factor equal to the ratio of the maximum absolute amplitude of the reference acoustic sensor waveform over the maximum absolute amplitude of the $n$'th acoustic sensor waveform. Synthetic tests (supplementary material, Figure S8) were performed to assess the resolution of the method using the OFAS array geometry presented above. A detailed description of the method is given in the supplementary material.

**4 Results**

4.1 Mechanical behavior of stick-slip instabilities

Stick-slip experiments presented in this study were performed at confining pressure $Pc$ ranging from 10 to 90 MPa. All experiments were conducted using a similar fault geometry and imposing a constant displacement rate resolved on the fault plane of around 1 $\mu m/s$. Figure 1a reports the evolution of both shear stress and fault slip with time for a stick-slip experiment at 60 MPa confining pressure. Increasing the axial stress leads first to the elastic increase of both shear stress and normal stress acting on the fault plane. Once the shear stress reaches a critical value $\tau_c$, corresponding to the critical strength of the fault, slip initiates leading to an abrupt stress release. The stress drop is proportional to the slip and both increase with the confining pressure. Regardless of the confining pressure, the system displays the same mechanical behavior. Figure 1b shows that slip increases linearly with the stress drop for all stick-slip experiments. The slope is equal to the stiffness of the whole system (machine and rock specimen). This has been observed in many other experiments on crustal rocks and can be explained by the increase of the normal stress on the fault with increasing in confining pressure, which enhances the strain energy stored in the medium during loading (Brace and Byerlee, 1966, Byerlee and Brace, 1968, Johnson et al., 1973, Johnson and Scholz, 1976, Passelègue et al., 2016).

4.2 Influence of rupture velocity and confining pressure on high-frequency radiation

The relation between the inverted rupture velocities, the stress drop and the confining pressure is shown in Figure 2a. Rupture velocities are normalized by the S-wave velocity of the medium, values under 0.92 correspond to sub-Rayleigh ruptures and values above 1 correspond to supershear ruptures. The overall trend of the rupture velocity is to increase with confinement and stress drop. For stress drops higher than 10 MPa, only supershear ruptures are observed. This was already well described by Passelègue et al. [2013] and can be understood in terms of the seismic ratio $S$ and the initial strength that precedes the rupture. $S$ controls the transition from sub-Rayleigh to supershear rupture (Andrews, 1976) and can be expressed as:

$$S = \frac{\tau_p - \tau_0}{\tau_0 - \tau_r}$$

where $\tau_p$, $\tau_0$ and $\tau_r$ are respectively the peak frictional stress, the initial stress and the residual frictional stress. Ruptures may transition from sub-Rayleigh to supershear velocity if the two conditions are satisfied: (i) the size of the fault is larger than the transition length from sub-Rayleigh to supershear rupture propagation $L_c$ which decreases with normal stress (ii) $S$ is smaller than $S_c$ (equal to 1.77 or 1.19 in 2D or 3D respectively).

In our experiments, the initial stress was always very close to peak frictional stress so that $S < S_c$ was always satisfied. However, estimates of $L_c$ at low confinement ($Pc \leq 20$ MPa) give values that are larger or of the same order of the size of our experimental fault, which explains why most of the ruptures were sub-Rayleigh at $Pc \leq 20$ MPa. Additional details are given in the supplementary material.

In Figure 2b the Fourier spectra that correspond to the last stick-slip event at each confining pressure are displayed (star symbols, Figure 2a). Directivity effects cannot be fully suppressed because our acoustic sensor network is not perfectly symmetric. Hence Fourier spectra were averaged over all acoustic sensors (i.e. from both AFAS and OFAS arrays) in order to minimize directivity bias. To compare the high-frequency content of the spectra, the latter have to scale at low frequency. As we expected the stress-drop to control the amplitude of low frequency waves, each spectrum is normalized by its corresponding stress-drop. We find a double correlation between the spectral amplitude of high-frequency radiation, the rupture velocity and the confining pressure. This is particularly well illustrated at the lowest confining pressure ($Pc = 10$ MPa), where stick-slip events ruptured at sub-Rayleigh velocity. The Fourier spectrum of these events is strongly depleted of high frequencies. In contrast, the effect of the confining pressure prevails over the effect of the rupture velocity, in the high-frequency radiation range, when comparing the spectra at $Pc = 20$ and 30 MPa ($Vr = 4500$ and 4100 m/s respectively). Similarly, the Fourier spectra at $Pc = 45$, 60 and 90 MPa which correspond to the highest rupture velocities ($Vr = 4900$, 5200, 4700 m/s respectively) are the most enhanced in high-frequency radiation. Note that at $Pc \geq$ 20 MPa, we consistently observe the emergence of two frequency bands. The first one is centered at 100 kHz and the second one lies between 400 and 800 kHz. In the following section, we show results of back-projection analysis applied to acoustic waveforms (i) bandpass filtered to 400-800 kHz and (ii) highpass filtered above 800 kHz.

### 4.3 Back-projection analysis during rupture propagation.

Unfiltered and band-pass filtered between 400 and 800 kHz OFAS waveforms are displayed in Figure 3. Waveforms are lined up with the first P-wave arrivals at each station. Only filtered waveforms were used for back-projection. We implicitly make the hypothesis that high-frequency sources are located on the fault plane. This assumption seems reasonable given that new fracture formations were never observed during any of the experiments performed for this study. Because our sensors are single components, we are not able to distinguish between P and S waves (and also surface waves and reverberations), which would make the back-projection results poorly resolved. As a consequence, the back-projection analysis are restrained to the beginning of the acoustic waveforms, i.e. before first S-wave arrivals at each stations (on average, 6 µs after first P-wave on the OFAS array). P-wave signals are back-projected on the fault plane by computing the coherency function over 2 µs time windows, with respect to the nucleation time. Figure 4 presents back-projection results in the 400-800 kHz frequency band (top) and above 800 kHz (bottom) for one event at $Pc$ = 90 MPa whose average rupture velocity was 5.1 km/s. The color-bar indicates the value of the coherency function normalized by its maximum value. The red star indicates the position of the nucleation and the black dashed line the theoretical position of the rupture front (at 1 µs for the 0-2 µs time window, at 2 µs for the 1-3 µs time window and so on) according to the estimated rupture velocity in section 4.2. In the supershear case, this theoretical rupture front is elliptical and propagates at constant velocities $C_s$ and $V_r$ along the ellipse's minor and major axes respectively, where $C_s$ and $V_r$ are the S-wave and in-plane rupture velocities (see supplementary material).

The 400-800 kHz frequency band (Figure 4 top) gives the clearest results. Throughout the rupture history, high-frequency energy sources are always localized behind the theoretical rupture front position. When rupture initiates (0-2 µs) high-frequency energy localizes slightly behind the nucleation and spreads over the width of the fault plane. At $t$ = 1-3 µs period, high-frequency energy starts to propagate consistently in the direction of the rupture front at relatively low speed and spreads over the entire width of the fault. The source of high-frequencies then accelerates (2-4 µs) along the fault plane until it roughly reaches the average rupture velocity (3-5 µs, 4-6 µs) while concentrating in the middle of the fault. Compared to the 400-800 kHz frequency band, back-projection images for high-frequency sources above 800 kHz (Figure 4 bottom) are less clear. When rupture initiates (0-2 µs), the maximum coherence is still focused close to the nucleation zone. It was also observed that the maximum coherence propagating consistently matched the theoretical rupture front (1-3 µs, 2-4 µs, 3-5 µs), although high-frequency energy was more diffuse and patchy. In contrast, between 4-6 µs, high-frequency energy starts to diffuse over the entire fault. Also, relative to high-frequency energy between 400-800 kHz, high-frequency energy above 800 kHz is always focused closer to the theoretical rupture front.

## 5 Discussion and conclusions

We summarize below the four key conclusions of this body of work.

High frequency radiation is related to stress and rupture velocity conditions: Observations of Fourier analysis (Figures 2a and 2b) have shown that high-frequency radiation is enhanced with both the stress conditions (i.e. normal stress acting on the fault) and the rupture velocity. This is consistent with seismological observations of mega-thrust subduction earthquakes where zones of high-frequency energy release correspond to deeper portions of the fault (Ishii et al., 2011). There are different ways to interpret these results. First, the increase of stress concentrations in the process zone with stress conditions and rupture velocity would likely enhance physical processes as off-fault damage (Thomas et al., 2017, Thomas and Bhat, 2018, Okubo et al. 2018) taking place in the vicinity of the rupture front leading to more radiated high-frequency energy. Also, as the rupture velocity increases, more abrupt acceleration/deceleration phases of the rupture front develop, leading to local slip accelerations which would enhance high-frequency radiation (Olson and Apsel, 1982, Hartzell and Heaton, 1983). Our laboratory observations may further our understanding of high-frequency radiation under controlled conditions.

High frequency radiation content depends on the speed regime: Here, we observed a net enhancement of high-frequency radiation when the rupture transitions from sub-Rayleigh regime to supershear regime (Figures 2a and 2b), in agreement with what has been proposed by previous studies (Bizzarri and Spudich, 2008, Vallèe et al., 2008). In order to investigate the consequences of supershear rupture velocities to high-frequency, we give an order magnitude estimate (Figure 2b) of the theoretical corner frequencies $f_c$ of far-field displacement spectrum for rupture velocities equal to $0.8*C_s$ (~ 2800 m/s) and to $1.4*C_s$ (~ 5000 m/s) based on the kinematic model for a circular crack of Sato and Hirasawa [1973] (see supplementary information for details). This agrees well with the observations for a sub-Rayleigh rupture but the model underestimates the corner frequency for the supershear case. This could be either because of model limitations or the fact that the geometric attenuation for supershear ruptures is significantly different (Dunham and Bhat, 2008).

Back-projection at laboratory scale provides new insights into earthquake processes: The fact that (i) we have been able to coherently back-propagate high-frequency energy at 400-800 kHz (ii) Fourier spectra show high-frequency asymptotes like $f^{-2}$ independent of the confining pressure (iii) the peak of energy at 100 kHz is absent at low confinement ($Pc$ = 10 MPa) strongly suggest that the information contained in the spectra is linked to the source. Thus, back-projection analysis (Figure 4) can provide new insights on the radiation of high-frequency waves and rupture processes. We carefully ensured that the back-projection results are reliable and are not manifestations of system noise (see supplementary material for details). The most robust and interpretable back-projection result obtained was in the 400–800 kHz frequency band (Figure 4 top). The correlation between the spatial and temporal evolution of high-frequency sources and the propagation of the rupture front provides concrete experimental evidence that high-frequency waves are concurrent with the propagation phase of the rupture front and that high-frequency radiation is emitted close to or behind the rupture tip. This result is in agreement with most of the studies that addressed the issue of high-frequency-radiation which proposed that high-frequency radiation is related to changes in rupture velocity due to fault stress or frictional heterogeneity, and predict high-frequency waves to be mainly generated in the vicinity of the rupture front (Madariaga, 1977, Madariaga, 1983, Haskell, 1964, Aki, 1967, Spudich and Frazer, 1984). Recent numerical studies (Thomas et al., 2017, Thomas and Bhat, 2018, Okubo et al., 2018) also demonstrated that high-frequency radiation was highly enhanced when co-seismic damage was implemented in their rupture propagation models. This is supported by microscopic analysis of the fault surface after stick-slip experiments under Scanning Electron Microscopy (see supplementary material), which revealed the presence of microcracks at the grain scale. Above 800 kHz (Figure 4, bottom), the back-projection results are less clear. It is not surprising given the fact that the signal to noise ratio is significantly lower relative to the 400-800 kHz frequency band and also that acoustic waves above 800 kHz are more sensitive to scattering effects due to small-scale heterogeneities. It might explain why, between 4-6 μs, high frequency energy diffuses over the entire fault. However, an observable feature is that high-frequency sources above 800 kHz (Figure 4 bottom) seem to localize slightly forward ahead of the one at 400-800 kHz. One hypothesis is that high-frequency radiation above 800 kHz highlights other physical processes. For instance, Doan and Gary [2009] suggested that grain pulverization and comminution and small-scale gouge particles production could produce high-frequency radiation. Such processes should indeed happen within the breakdown zone, very near the rupture front and should be followed by asperity melting (Passelègue et al, 2016, Aubry et al, 2018).

Back-projection method can approximate the geometry of high frequency sources: Finally, synthetic tests (Figure S8, supplementary material) demonstrated that the back-projection method can approximately image the high-frequency source geometry. Back-projection results at 400-800 kHz have shown that at the beginning of the rupture and during rupture propagation, high-frequency radiation is drawing a pattern that is spread over almost the entire width of the fault and that is linear along the width of the fault, although it is less noticeable between 4 μs and 6 μs. However, because acoustic recordings have been aligned to the nucleation zone, the cross-correlation procedure is expected to be less efficient as the source is moving away from the nucleation. This could explain why the initial pattern is not preserved and is concentrated in the middle of the fault with time. Under the assumption that high-frequency sources are representative of the shape of the rupture front, the observations do not match with what would be expected for an elliptical crack in an infinite medium but that of a rupture front strongly interacting with a free surface (Fukuyama et al., 2018, Passelègue et al., 2016).

This study has shown that back-projection analysis at the laboratory scale could be of relevance to understand the nucleation and propagation dynamics of earthquakes. In the future, the combined use of additional phases (S-waves, surface waves, reflected waves) and the deconvolution of acoustic recordings from Green's function describing the medium should help to get a more detailed and complete description of the source.

## Acknowledgments

We thank Professor Miaki Ishii (Harvard University, USA) for her introduction to the use of the back-projection technique. Authors would also like to thank Damien Deldicque (ENS Paris, France) for his help in producing fault surface images under Scanning Electron Microscopy. Authors would also like to thank the anonymous reviewers who significantly participated in improving the quality of our study. This work was funded by the European Research Council grant REALISM (2016-grant 681346). The authors declare that they have no competing financial interests. All data are available at : https://github.com/samsonmarty/high-frequency-radiation-during-laboratory-earthquakes.

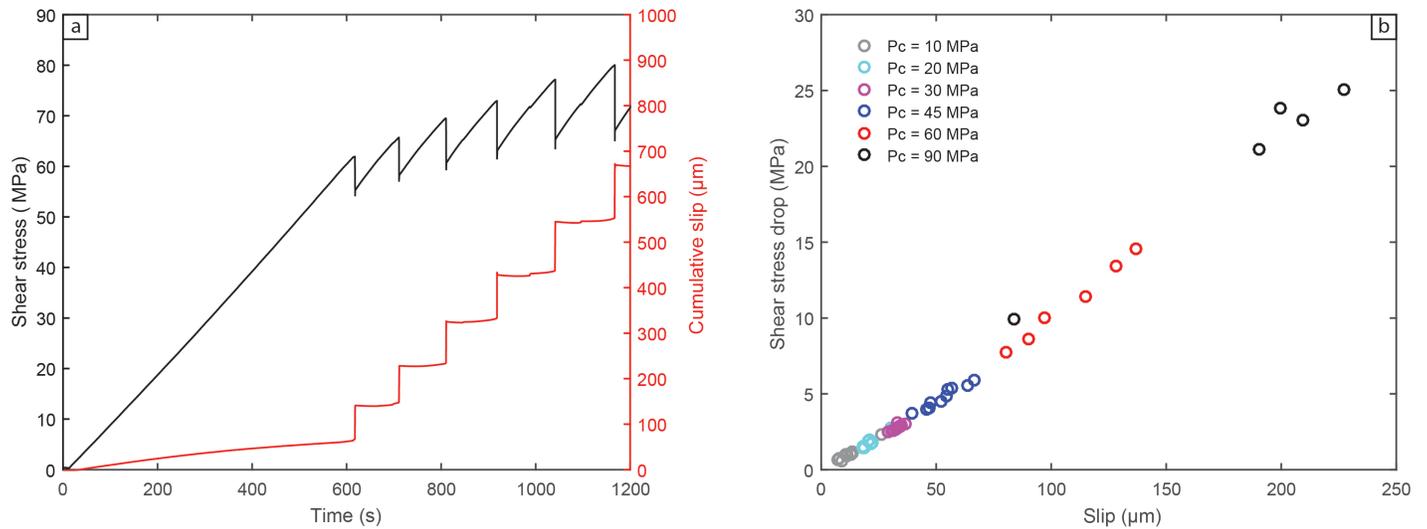

**Figure 1.** a) Evolution of shear stress and slip versus time at $Pc = 60$ MPa. When the shear stress on the frictional interface exceeds the fault strength the stored elastic energy is suddenly released by seismic slip. The cumulative slip remains constant during loading because it is corrected from the elastic part of the deformation (sample and apparatus). (b) Relationship between shear stress drop and slip for all experiments. The ratio between the stress drop and the slip is preserved (higher the stress drop, higher the amount of slip) and is equal to the stiffness of the whole system (sample and apparatus).

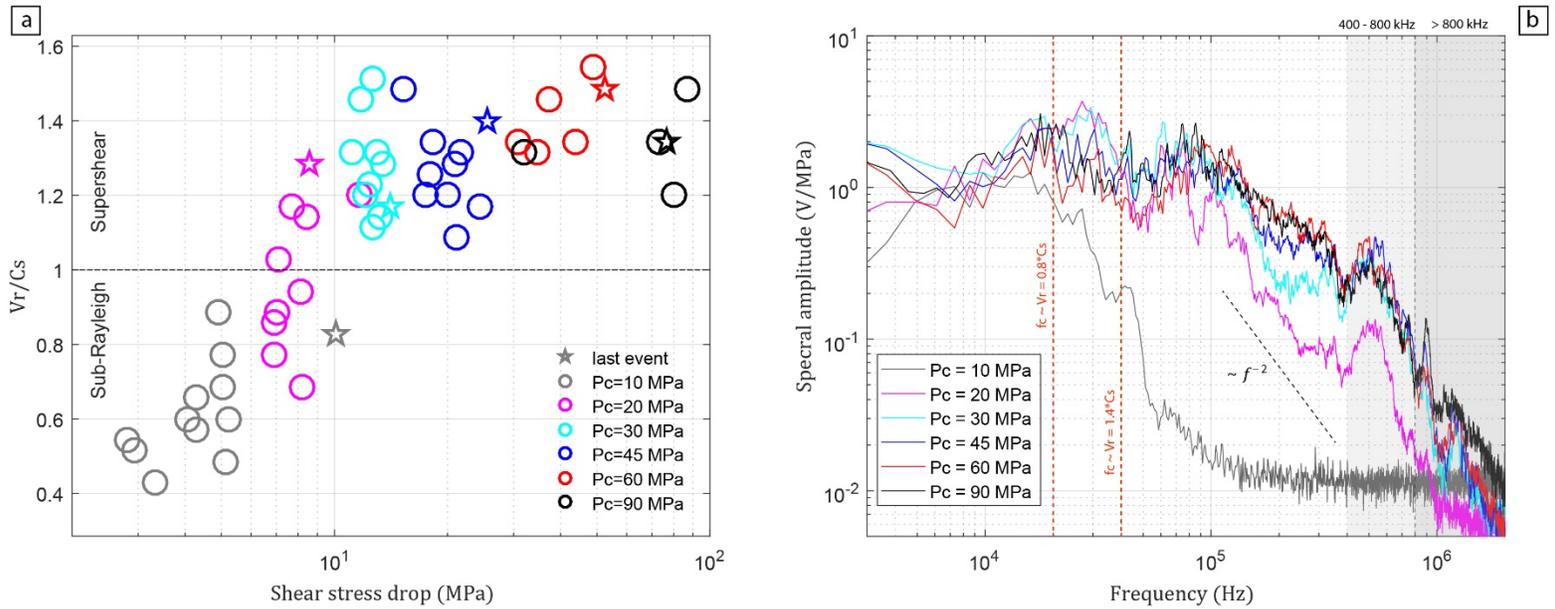

**Figure 2.** (a) Rupture velocity obtained by inversion as a function of static shear stress drop. Rupture velocities are normalized by the shear wave velocity, values higher than 1 correspond to supershear velocities and lower than 0.92 to sub-Rayleigh velocities. Stars indicate stick-slip events whose Fourier spectra are displayed in Figure 2b. (b) Fourier spectra of the last stick-slip event during stick-slip experiments at varying confining pressures. Fourier spectra are averaged using both AFAS and OFAS arrays, and normalized by their respective stress-drop. The gray shaded areas indicate frequency bands used for the back-projection analysis.

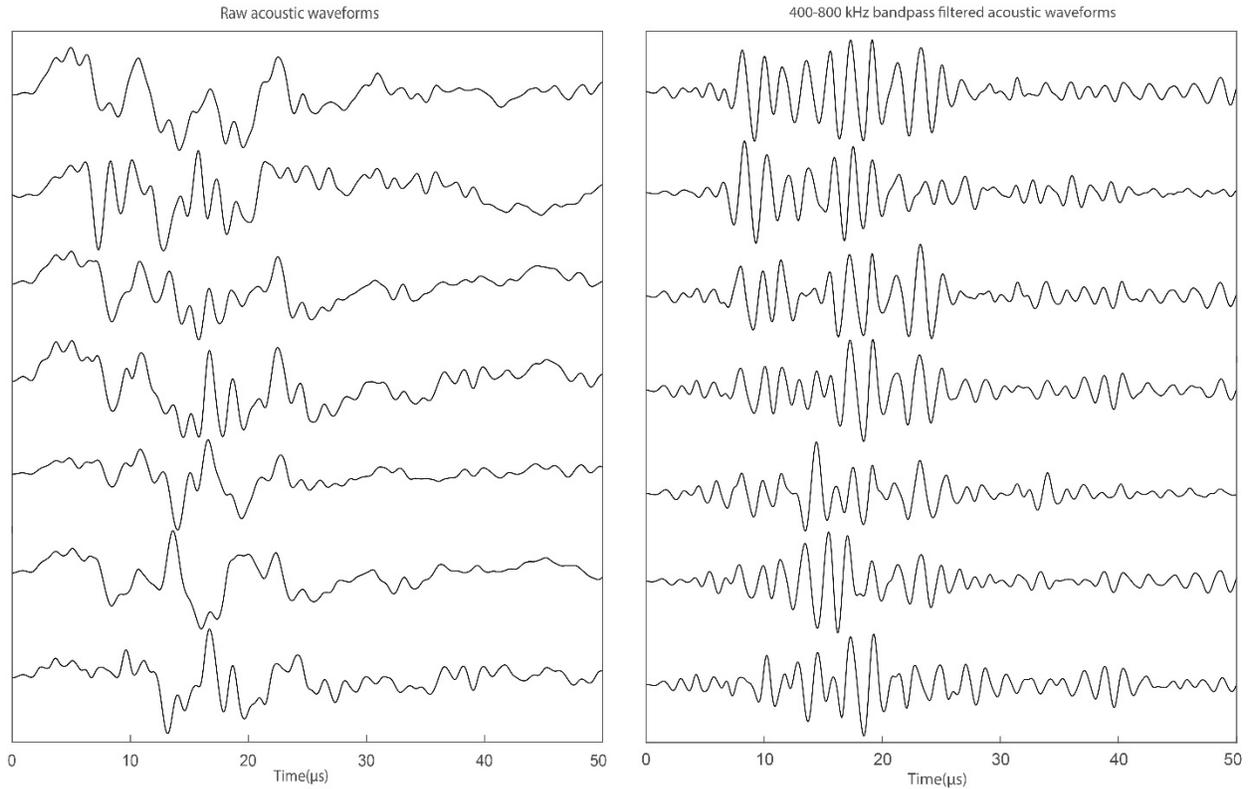

**Figure 3.** Example of acoustic waveforms used for the back-projection analysis: raw acoustic waveforms (left) and band-pass (400-800 KHz) acoustic waveforms (right). In both cases waveforms are aligned on the first P-wave arrivals and are normalized by their maximum amplitudes.

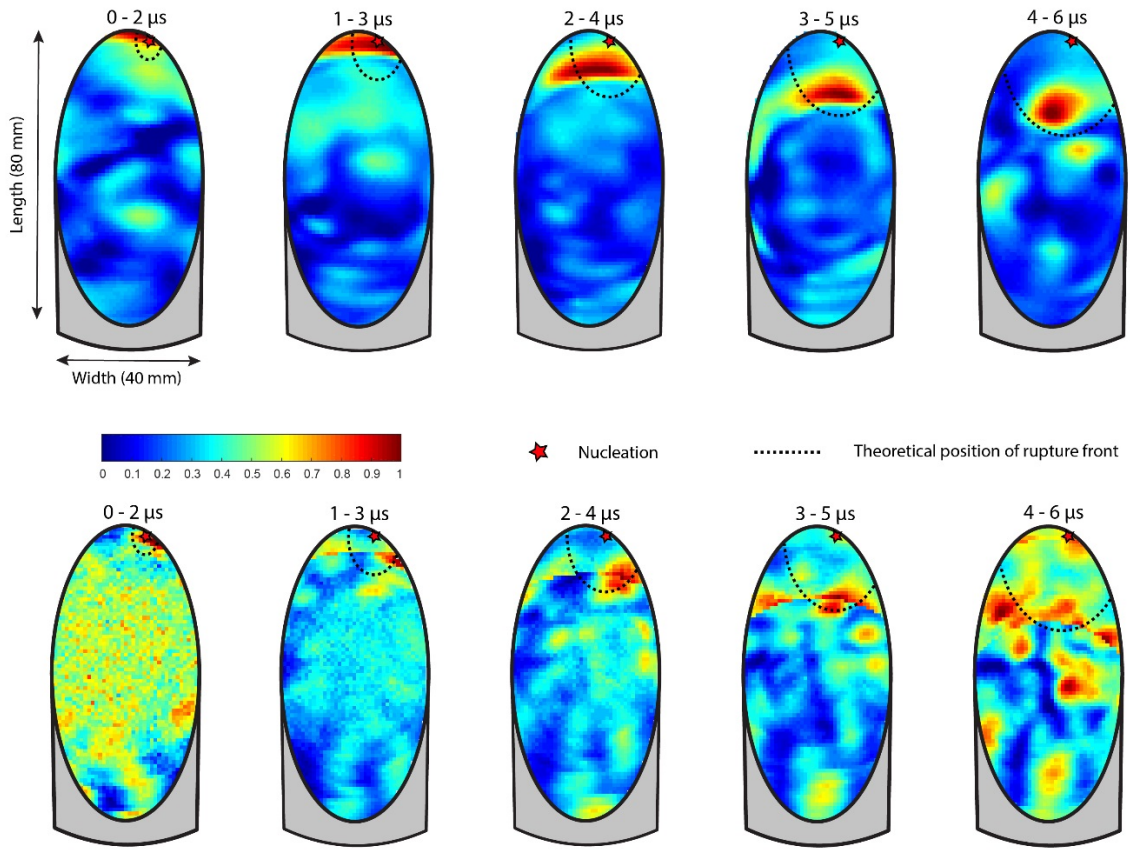

**Figure 4.** Snapshots of back-projection results for one stick-slip event at $Pc$ = 90 MPa from OFAS waveforms bandpass filtered to 400-800 kHz (top) and highpass filtered above 800 kHz (bottom). The colorbar represents the value of the coherency function on the fault plane. The time is relative to the onset of the nucleation. The red star indicates the nucleation location and the black dashed line indicates the rupture front theoretical position estimated from the average rupture velocity Vr obtained by inversion, here equal to 5.1 km/s.